\documentclass[a4paper]{jpconf}
\usepackage{graphicx}
\usepackage{amstext,amsmath,amssymb,amsfonts,bbm}
\def\be{\begin{equation}}
\def\ee{\end{equation}}
\def\bes{\begin{eqnarray}}
\def\ees{\end{eqnarray}}

\def\6{\langle}
\def\9{\rangle}

\def\1{{\mathbbm 1}}

\begin{document}
\title{Measurements of non local weak values}

\author{Aharon Brodutch$^{1,2}$ and Lev Vaidman$^1$}

\address{$^1$Raymond and Beverly Sackler School of Physics and Astronomy, Tel Aviv University, Tel Aviv, Israel

$^2$Center for Quantum Computer Technology, Faculty of Science, Macquarie University, Sydney NSW 2113, Australia
}

\ead{abrodutc@ics.mq.edu.au}

\begin{abstract}
Some recent attempts at measuring non local weak values via local measurements are discussed
and shown to be less robust than standard weak measurements. A method for measuring some non
local weak values via non local measurements (non local weak measurements) is introduced. The
meaning of  {\it non local weak values} is discussed.

\end{abstract}

\section{INTRODUCTION}
The concept of {\it observable} or {\it variable} in quantum
mechanics is not as simple and intuitive as in classical physics and it
requires definition. While in classical physics  every observable has a value for any state of the
system (which might be known or not known to the observer), in quantum mechanics we cannot associate a
value for every observable. Instead, a measurement procedure is
defined and every variable is associated with a set of  {\it
eigenvalues}, the possible outcomes of its measurement. Every state
of a quantum system is associated with probability distributions for
outcomes of measurements of every variable. It might happen that
for some variables the probability distribution is singular, i.e., a
particular eigenvalue is obtained with certainty and in this case,
this eigenvalue can be associated with the variable as in the classical
case. Otherwise, a statistical expectation value can be associated
with a variable.

The definition of a quantum measurement procedure is, therefore,
crucial for the concept  of the  value of a variable in quantum
mechanics. The standard model of quantum measurement is the Von
Neumann procedure \cite{vonNeumann} which consists of a  short interaction
between the system and a measuring device. The quantum measurement of a
variable $O$ is described by the Hamiltonian :
\begin{equation}
H=g(t)PO,\label{neumann}
\end{equation}
 where $P$ is the momentum conjugate to the pointer variable of the measuring device $Q$,
and the normalized coupling function $g(t)$ specifies the time of
the measurement interaction. The outcome of the measurement is the
shift of the pointer variable during the interaction. In an ideal
measurement the function $g(t)$ is nonzero only during a very short
period of time, and the free Hamiltonian during this period of time
can be neglected.

A discussion of ideal and non ideal measurements requires a few definitions. An {\it ideal,  nondemilition, instantaneous}, measurement is defined as a Von Neumann measurement that leaves the state of a system, which initially was in an eigenstate of the measured observable,  unchanged. A faithful demolition measurement is one that gives the required result but disturbs the system by changing its state. Nonlocality too needs definition in this context. A nonlocal system is one that has two or more parts placed in separate locations. A local system is one that exists in a single location. Local interactions are those that can be created between two systems in the same place.

Formally, in non relativistic quantum mechanics one can consider any
Hamiltonian and thus one can measure any variable. However,
relativistic quantum mechanics limits us to local interactions and
thus we cannot construct the Hamiltonian (\ref{neumann}) for a variable
$O$ related to a composite system with parts placed in separate
locations. Nevertheless, measurements of some nonlocal variables are
possible \cite{AAVnonlocal} in the sense that at the end of an instantaneous
measurement procedure the composite system ends up in an eigenstate
of this variable and the information about the eigenstate is written
down (although in separate locations). We call this procedure a {\it
nonlocal measurement}.

For local measurements we can consider the standard Von Neumann
procedure with weakened coupling. In such measurements the pointer
does not point sharply to  zero before the interaction, and  it does
not point sharply to an eigenvalue after the interaction. The
probability distribution of the pointer after the interaction points
at the expectation value. Such measurements are called {\it weak
measurements}. Particularly interesting are weak measurements
performed on pre and post-selected quantum systems. For a quantum
system pre-selected in a state $\vert\Psi\rangle$ and post-selected
in a state $\vert\Phi\rangle$  the probability distribution of the
pointer variable points to (a real part of) the {\it weak value}
\cite{timeinQM}
 \begin{equation}
O_{w}\equiv\frac{\langle{\Phi}\vert
O\vert\Psi\rangle}{\langle{\Phi}\vert{\Psi}\rangle}.
\label{wv}\end{equation}

In some cases weak values might be much larger than the eigenvalues of $O$. This
amplification is sometimes called the Aharonov-Albert-Vaidman (AAV)
effect.

The simple formula above defines the weak value of any variable, including non local variables.  In this paper we consider a possibility of combining the ideas of weak and nonlocal measurements to measure such non local weak values. This work was inspired by recently introduced concept of \textit{joint weak measurements} \cite{ReschSteinberg}. In
this procedure a new analysis of the readings on local measuring devices which
performed  local weak measurements allows  to calculate the weak
values of the product of two variables related to separate
parts of the system. We analyze this procedure and show that  it lacks some of the fundamental features of weak measurements and requires much larger resources than standard weak measurements.  We show that some non local weak values can be measured using \textit{non local weak measurements} and introduce a method for making such measurements. We argue that only  values (local or non local) that can be measured directly using a weak measurement can be thought of as weak values, while those that are measured indirectly are just the result of a calculation leading to the result of (\ref{wv}).

\section{Nonlocal Measurements}

The special theory of relativity limits us  to local interaction,
so if $A$ is a nonlocal variable related to separate locations, the
interaction described by (\ref{neumann}) does not exist in nature.
Nevertheless there are some nonlocal measurements which can be performed.
A composite measuring device is used, with parts near every location
of the variable $A$.

For example, a nonlocal variable $A=\sum B_{i}$
where each variable $B_{i}$ is related to location $i$ can be measured
with the Hamiltonian \begin{equation}
H=g(t)\sum P_{i}B_{i},\label{neumannN}\end{equation}
 where $P_{i}$ is the momentum conjugate to the pointer variables
$Q_{i}$ located in location $i$.

If initially all pointer variables
are well localized around zero, we can learn the outcome of a measurement
of $A$ from the sum of readings of all local measuring devices $A=\sum Q_{i}$.
This procedure is a faithful measurement of $A$, but it is not an
ideal measurement of $A$. In an ideal measurement a system initially  in an eigenstate of  $A$  should remain unchanged after the measurement. This is usually not the case for the method described above. For example, if our system consists of two spin$-\frac{1}{2}$
particles in a singlet state\\
$\frac{1}{\sqrt{2}}(|\uparrow\rangle|\downarrow\rangle-|\downarrow\rangle|\uparrow\rangle)$
and $A=\sigma_{1z}+\sigma_{1z}$, then we will learn from our
measurement that $A=Q_{1}+Q_{2}=0$, but the singlet state
will be changed either to $|\uparrow\rangle|\downarrow\rangle$ or to
$|\downarrow\rangle|\uparrow\rangle$.

In order to perform the nonlocal measurement we should do the following \cite{AAVnonlocal}. Instead of localizing all pointers $Q_{i}$ we start with the following entangled state of the measuring device: \begin{equation} |\Psi\rangle_{in}^{MD}=|\sum
Q_{i}=0,P_{1}=P_{2}=...=P_{N}\rangle\label{nlsumpointer}\end{equation}
 After the interaction, we can read the value of $A$ as before: $A=\sum Q_{i}$,
but now the eigenstates of $A$ remain unchanged.

It has been shown \cite{AAVnonlocal} that beyond the measurement of a
sum of local variables separated in space one can also perform a
measurement of a modular sum of local variables. But there are
nonlocal variables which cannot be measured in a non-demolition way
\cite{AAVnonlocal}. In particular, consider a product,
$\sigma_{z}^{A}\sigma_{z}^{B}$, of spin components of two separate
spin-1 particles, one in Alice's hand and another in Bob's hand.We now prove that the possibility of an ideal non-demolition measurement of that variable contradicts causality, and thus, such a measurement does
not exist.

Before time $t=0$ we prepare the two particles in the
initial state
\begin{equation}
|\Psi\rangle_{in}=\frac{1}{\sqrt{2}}(|-1\rangle_{A}+|0\rangle_{A})~|0\rangle_{B}\label{insta}\end{equation}
We assume that at time $t=0$ somebody performs a non-demolition measurement
of $\sigma_{z}^{A}\sigma_{z}^{B}$. Immediately after $t=0$ Alice performs
her projective local measurement of the state $\frac{1}{\sqrt{2}}(|-1\rangle_{A}+|0\rangle_{A})$.
Bob, who has  access to particle $B$ can send a superluminal signal
to Alice in the following way. Just before $t=0$ he decides to change
the state of his spin to $|1\rangle_{B}$ or to leave it as it is
$|0\rangle_{B}$. If he decides to do nothing, then a nonlocal measurement
of $\sigma_{z}^{A}\sigma_{z}^{B}$ will not change the state of the particles
because state (\ref{insta}) is an eigenstate of $\sigma_{z}^{A}\sigma_{z}^{B}$.
Therefore, Alice, in her local projective measurement will find the
state $\frac{1}{\sqrt{2}}(|-1\rangle_{A}+|0\rangle_{A})$ with certainty.
However, if Bob decides to change the state of his spin to $|1\rangle_{B}$
the initial state before the nonlocal measurement at $t=0$ will be
changed to \begin{equation}
|\Psi'\rangle_{in}=\frac{1}{\sqrt{2}}(|-1\rangle_{A}+|0\rangle_{A})~|1\rangle_{B}\label{insta1a}\end{equation}
 It is not an eigenstate of $\sigma_{z}^{A}\sigma_{z}^{B}$ and thus after
the measurement we will end up with equal probability with the state
$|-1\rangle_{A}~|1\rangle_{B}$ or $|0\rangle_{A}~|1\rangle_{B}$.
In both cases the probability of obtaining a positive outcome in Alice's's
projective measurement on the state $\frac{1}{\sqrt{2}}(|-1\rangle_{A}+|0\rangle_{A})$
is just one half. Instantaneous change of probability of a measurement
performed by Alice breaks causality, therefore instantaneous measurement
of $\sigma_{z}^{A}\sigma_{z}^{B}$ is impossible.

Note that not any product is unmeasurable. If instead of two spin-1
particles we consider two spin-$\frac{1}{2}$ particles then the product
$\sigma_{z}^{A}\sigma_{z}^{B}$ is measurable. Indeed, we can express this
product as a modular sum: $\sigma_{z}^{A}\sigma_{z}^{B}=(\sigma_{z}^{A}\sigma_{z}^{B})mod4-1$
and every modular sum is measurable.

If we relax the requirement of an ideal measurement that it should
be a non-demolition measurement, (which is a very rare property of
real quantum measurements) and require only that it gives us a
faithful result, then, conceptually, there are no constraints on
measuring nonlocal variables. Given a large enough resource of entanglement
we can {}``teleport'' the quantum states of all separate parts of
the system to one location and perform a measurement of any variable
\cite{vaidman-2001}. This is not a real teleportation which requires
sending classical bits, this procedure can be performed
instantaneously. Of course, the result of measurement can only be read
later, when the results of the local measurements will be brought
together. If the system is in addition pre and post-selected, the
procedure should be slightly modified, although (somewhat
surprisingly) we need not add a lot of entanglement resources
\cite{nonlocaltimesymmetric}.

\section{WEAK MEASUREMENTS}

 A weak measurement is a standard Von Neumann measurement with weakened interaction. One of the ways to weaken the interaction is to prepare the state of the measuring device in such a way that $P$ is very small. A good model of weak measurement is given by the coupling (\ref{neumann})
with the initial state of the pointer variable a given by a Gaussian centered
around zero:
 \begin{equation}
\Psi_{in}^{MD}(Q)=(\Delta^{2}\pi)^{-1/4}e^{-{{Q^{2}}/{4\Delta^{2}}}}.
\label{md-in}\end{equation}
with the position uncertainty  $\Delta$  large ensuring a small $P$.  We will henceforth use this model to describe weak measurements.

Weak values might lie very far from the range of the eigenvalues.
For example, a spin half particle prepared in the $x$ direction and
post-selected in an almost orthogonal state\\
$\frac{1}{2}(\cos(\frac{\pi}{4}+\epsilon)|\uparrow\rangle-\sin(\frac{\pi}{4}+\epsilon)|\uparrow\rangle$
will yield a very large weak value for $\sigma_{z}$ \cite{spin100}. Even in these cases there is an obvious shift of the pointer variable by the weak value (\ref{wv}). This shift gives weak values their significance \cite{weakreality}.

While the meaning of weak values remains controversial \cite{Replyleggett,Replyperes,Replyaharonov}, the justification of considering  weak values as a description of the pre and post-selected
quantum systems relies on the universality of the influence of the
coupling to a variable in the limit of its weakness. The pointer
variable prepared in a natural way (see Jozsa for some limitations
\cite{complexweak}) shifts due to a weak measurement coupling as if it
were coupled to a classical variable with the value equal to the weak
value.

 There have been numerous experiments showing weak values
\cite{Ex1,Ex2,Ex3,Ex4,Ex5}, mostly of photon polarization and the
AAV effect has been well confirmed. Hosten and Kwiat
\cite{HK} applied weak measurement procedure for measuring spin Hall
effect in light. This effect is so tiny that it can not be observed
without the amplification.

We must not forget that  there is a method for performing any (demolition) measurement \cite{vaidman-2001}. Thus, given large enough ensemble we can,
in particular, measure the two-state vector at the time of the measurement.
This is a complete description of a pre and post-selected quantum
system, so it allows to calculate any function of the pre and post-selected
state, and among others, the weak value or rather the result of (\ref{wv}).

\section{LOCAL AND NONLOCAL MEASUREMENTS OF THE SUM OF LOCAL VARIABLES}

We will start with the simplest example of non local variables, the sum of two local variables.  Using the expression (\ref{wv}) we get
\begin{equation}
(A+B)_{w}=\frac{\langle{\Phi}\vert
A+B\vert\Psi\rangle}{\langle{\Phi}\vert{\Psi}\rangle}\end{equation}
which in turn gives us the simple relation:
\begin{equation}
(A+B)_{w}=A_{w}+B_{w}
\label{weaksum}\end{equation}
 Thus, in order to find the value of $(A+B)_{w}$ we can measure
 $A_{w}$ and $B_{w}$ locally and just  add the two numbers. This fact is
somewhat surprising because the analogous relation for an expectation
values of the measuring devices performing strong measurements on pre and post-selected systems (which we signify as $\langle ~~~~ \rangle _{\Phi \Psi}$) does not hold. In general
 \begin{equation}
\langle A+B\rangle_{\Phi \Psi}\neq\langle A\rangle_{\Phi \Psi} ~+~\langle B\rangle_{\Phi \Psi}\label{strongsum}
\end{equation}
 Consider the following example . At time $t_{1}$ two spin$-\frac{1}{2}$
particles are prepared in a state
\begin{equation}
|\Psi\rangle=\sqrt{\frac{1}{2+\epsilon^2}}(|\uparrow\rangle_{A}|\downarrow\rangle_{B}+|\downarrow\rangle_{A}|\uparrow\rangle_{B})+\epsilon|\uparrow\rangle_{A}|\uparrow\rangle_{B}
\label{insta1}\end{equation}
 Later, at time $t_{2}$, the particles are found in a state
\begin{equation}
|\Phi\rangle=\sqrt{\frac{1}{2+\epsilon^2}}(|\uparrow\rangle_{A}|\downarrow\rangle_{B}-|\downarrow\rangle_{A}|\uparrow\rangle_{B})+\epsilon|\uparrow\rangle_{A}|\uparrow\rangle_{B}
\label{finsta1}\end{equation}
We can use the
Aharonov-Bergmann-Lebowitz (ABL)\cite{ABL1} formula for calculating
the probabilities for the outcomes of strong intermediate measurements of a variable $C$ given the pre-selection
$|\Psi\rangle$ and post-selection $|\Phi\rangle$
\begin{equation}
{\rm Prob}(c_{n})=\frac{{|\langle\Phi|{\bf P}_{C=c_{n}}|\Psi\rangle|^{2}}}{{\sum_{j}|\langle\Phi|{\bf P}_{C=c_{j}}|\Psi\rangle|^{2}}}\label{ABL}\end{equation}
where $P_{c=c_{j}}$ is a projection operator for the eigenstate(s)
with the eigenvalue $c_{j}$. Thus, for an ideal measurement of the nonlocal variable
$\sigma_{z}^{A}+\sigma_{z}^{B}$ we obtain:
\begin{eqnarray}
\langle\sigma_{z}^{A}+\sigma_{z}^{B}\rangle_{\Phi \Psi} & = & 2p(\uparrow\uparrow)+0p(\sigma_{z}^{A}+\sigma_{z}^{B}=0)-2p(\downarrow\downarrow)=
\label{strongsumsasb}\end{eqnarray}

\[
\frac{2|\langle\Phi|{\bf P}_{\uparrow\uparrow}|\Psi\rangle|^{2}-2|\langle\Phi|{\bf P}_{\downarrow\downarrow}|\Psi\rangle|^{2}}{|\langle\Phi|{\bf P}_{\uparrow\uparrow}|\Psi\rangle|^{2}+|\langle\Phi|{\bf P}_{\downarrow\downarrow}|\Psi\rangle|^{2}+|\langle\Phi|{\bf P}_{\sigma_{z}^{A}+\sigma_{z}^{B}=0}|\Psi\rangle|^{2}}=
\frac{2|\epsilon^{2}\frac{1}{2+\epsilon^2}|^{2}}{|\epsilon^{2}\frac{1}{2+\epsilon^2}]|+|0|^{2}+|0|^{2}}=2.\]
For an ideal measurement of a local variable $\sigma_{z}^{A}$, given that
it is the only intermediate measurement that has been performed we
have:

\begin{equation}
\langle\sigma_{z}^{A}\rangle_{\Phi \Psi}=p(\uparrow)-p(\downarrow)=\frac{|\langle\Phi|P_{\sigma_{z}^{A}=\uparrow}|\Psi\rangle|^{2}-|\langle\Phi|P_{\sigma_{z}^{A}=\downarrow}|\Psi\rangle|^{2}}{|\langle\Phi|P_{\sigma_{z}^{A}=\uparrow}|\Psi\rangle|^{2}+|\langle\Phi|P_{\sigma_{z}^{A}=\downarrow}|\Psi\rangle|^{2}}=
\label{strongsa}\end{equation}
\[
=\frac{|(\epsilon^{2}+1)\frac{1}{2+\epsilon^2}|^{2}-|(-1)\frac{1}{2+\epsilon^2}|^{2}}{|(\epsilon^{2}+1)\frac{1}{2+\epsilon^2}|^{2}+|(-1)\frac{1}{2+\epsilon^2}|^{2}}=\frac{2\epsilon^{2}+\epsilon^{4}}{{2+\epsilon^{4}+2\epsilon^{2}}}\]
 The expectation value of $\sigma_{z}^{B}$ measured alone is \begin{equation}
\langle\sigma_{z}^{B}\rangle_{\Phi \Psi}=p(\uparrow)-p(\downarrow)=\frac{|\langle\Phi|P_{\sigma_{z}^{B}=\uparrow}|\Psi\rangle|^{2}-|\langle\Phi|P_{\sigma_{z}^{B}=\downarrow}|\Psi\rangle|^{2}}{|\langle\Phi|P_{\sigma_{z}^{B}=\uparrow}|\Psi\rangle|^{2}+|\langle\Phi|P_{\sigma_{z}^{B}=\downarrow}|\Psi\rangle|^{2}}=\label{strongsb}\end{equation}
\[
\frac{|(\epsilon^{2}-1)\frac{1}{2+\epsilon^2}|^{2}-|(+1)\frac{1}{2+\epsilon^2}|^{2}}{|(\epsilon^{2}+1)\frac{1}{2+\epsilon^2}|^{2}+|(+1)\frac{1}{2+\epsilon^2}|^{2}}=\frac{-2\epsilon^{2}+\epsilon^{4}}{{2+\epsilon^{4}-2\epsilon^{2}}}\]
 It is easy to see that the expectation value of the nonlocal variable
(\ref{strongsumsasb}) is very different from the sum of (\ref{strongsa})
and (\ref{strongsb}). However, it is more reasonable to compare (\ref{strongsumsasb})
with the sum of expectation values of the outcomes of local measurements
of $\sigma_{z}^{A}$ and $\sigma_{z}^{B}$ performed simultaneously.
This corresponds to the measurement of a variable with non-degenerate eigenstates
$|\uparrow\rangle_{A}|\downarrow\rangle_{B},|\downarrow\rangle_{A}|\uparrow\rangle_{B},|\uparrow\rangle_{A}|\uparrow\rangle_{B},|\downarrow\rangle_{A}|\downarrow\rangle_{B}$.
In this case we have
\begin{eqnarray}
&\langle\{\sigma_{z}^{A}\}+\{\sigma_{z}^{B}\}\rangle_{\Phi \Psi}=2p(\uparrow\uparrow)+0p(\uparrow\downarrow)+0p(\downarrow\uparrow)=\frac{2|\langle\Phi|{\bf
P}_{\uparrow\uparrow}|\Psi\rangle|^{2}}{|\langle\Phi|{\bf
P}_{\uparrow\uparrow}|\Psi\rangle|^{2}+|\langle\Phi|{\bf
P}_{\uparrow\downarrow}|\Psi\rangle|^{2}+|\langle\Phi|{\bf
P}_{\downarrow\uparrow}|\Psi\rangle|^{2}}=\label{strongsumsa+sb}&\end{eqnarray}
\[
\frac{2|\epsilon^{2}\frac{1}{2+\epsilon^2}|^{2}}{|\epsilon^{2}\frac{1}{2+\epsilon^2}|^{2}+|\frac{1}{2+\epsilon^2}|^{2}+|(-1)\frac{1}{2+\epsilon^2}|^{2}}=\frac{2\epsilon^{2}}{{2+\epsilon^{4}}}\]
the brackets $\{\}$ signify a separate measurement of each observable.

We see that expectation value of the strong measurements of the sum
of two variables related to separate parts is not equal to the sum
of the expectation values of local measurements, even if they are performed simultaneously.
 \begin{equation}
\langle\sigma_{z}^{A}+\sigma_{z}^{B}\rangle_{\Phi \Psi}\ne\langle\sigma_{z}^{A}\rangle_{\Phi \Psi}+\langle\sigma_{z}^{B}\rangle_{\Phi \Psi}\ne\langle\{\sigma_{z}^{A}\}+\{\sigma_{z}^{B}\}\rangle_{\Phi \Psi}\label{strongsumsep}
\end{equation}

 For weak measurements, however, the equality (\ref{weaksum})  holds, both for separate
and joint local measurements.
If we assume the existence of nonlocal interactions, then we can directly
couple to the sum $\sigma_{z}^{A}+\sigma_{z}^{B}$. \begin{equation}
H=g(t)P(\sigma_{z}^{A}+\sigma_{z}^{B}).\label{neumannNL}\end{equation}
 Since a strong measurement yields the outcome 2 with certainty, the
weak value should  also equal 2 according to the theorem proved in
\cite{AV91}. In this particular example, the coupling need not be
weak to find the weak value 2. Since there are no nonlocal interactions in
nature we have to apply local weak measurements of $\sigma_{z}^{A}$
and $\sigma_{z}^{B}$
\begin{equation}
H=g(t)(P_{A}\sigma_{z}^{A}+P_{B}\sigma_{z}^{B})\label{neumannL}
\end{equation}
and add the outcomes.
We use a model in which the initial wave functions
of the measuring devices are Gaussians around zero
 \begin{equation}
\Psi_{in}^{MD}(Q_{A},Q_{B})=Ne^{\frac{{-Q_{A}^{2}}}{2\Delta^{2}}}e^{\frac{{-Q_{B}^{2}}}{2\Delta^{2}}}\label{ini}
\end{equation}
The measurement interaction leads to a shift of the
pointer wave function by the eigenvalue of $\sigma_{z}$ so that after the measurement, the measuring device will be described by the state
\begin{eqnarray}
\nonumber
 &{\Psi_{fin}^{MD}(Q_{A},Q_{B})=}&\\&
\Psi_{in}^{MD}(Q_{A}-1,Q_{B}+1)-\Psi_{in}^{MD}(Q_{A}+1,Q_{B}-1))+\epsilon^{2}\Psi_{in}^{MD}(Q_{A}-1,Q_{B}-1)=
&\\\nonumber
&Ne^{\frac{{-(Q_{A}^{2}+Q_{B}^{2})}}{2\Delta^{2}}}[e^{\frac{{-(-2Q_{A}+2Q_{B}+2)}}{2\Delta^{2}}-}
-e^{\frac{{-(2Q_{A}-2Q_{B}+2)}}{2\Delta^{2}}}+\epsilon^{2}e^{\frac{{-(-2Q_{A}-2Q_{B}+2)}}{2\Delta^{2}}}]&\nonumber
\end{eqnarray}
with $N$ being a normalization constant.

The expectation value of the measurement outcome is:
\begin{eqnarray}
\nonumber
&\langle Q_{A}+Q_{B}\rangle=\int(Q_{A}+Q_{B})|\Psi_{fin}^{MD}(Q_{A},Q_{B})|^{2}dQ_{A}dQ_{B}=&\\
&\frac{2\epsilon^{4}}{\epsilon^{4}+(2-2e^{\frac{-2}{\Delta^{2}}})}\approx
 2-\frac{8}{\Delta^{2}\epsilon^{4}}& \label{qa+qb}
\end{eqnarray}
for large $\Delta$. The statistical measurement error is the width $\Delta$. At the limit of weak measurements, i.e. large $\Delta$ it indeed yields the weak value  $(\sigma_{z}^{A}+\sigma_{z}^{B})_{w}=2$. However, we can go
close to the weak value only for very large $\Delta$, (very
weak measurement) and thus very large uncertainty in the final reading.
For example, in the case of  $\epsilon=0.1$ we will need to have $\Delta>600$
in order to measure the weak value with a deviation of 10\%.
 Of course, if we make the same
measurement on a large ensemble the statistical error will be made
smaller according to $
\Delta_{n}=\frac{\Delta}{\sqrt{n}}$
requiring us to use an ensemble of about $3.6\times10^{5}$ such pre
and post selected systems just to be within the right order of magnitude.
This number will be increased if we want to make the deviation or the
statistical error smaller. Since for a true nonlocal measurement we
can get the required result at the strong limit, we only need one
such system to get the correct expectation value.
We see that  local measurements allow us to find nonlocal weak
values, but it is a very inefficient procedure. The nonlocal weak value is the result of a calculation made on the readings of two pointer variables rather than the direct result of the reading of a single pointer variable.

Let us now try  to combine the techniques of nonlocal measurements based on a measuring
device with entangled parts and local interactions, with the weak
measurement techniques. The measuring device in an ideal strong
nonlocal measurement has an initial state (\ref{nlsumpointer}). To make it weak we have to prepare the conjugate momenta to be centered around $0$ which requires 
the pointer (centered around $Q_{A}+Q_{B}=0$) to have a large uncertainty.  In this case,
the initial state of the measuring device (using the usual Gaussian model) will be
\begin{equation}
\Psi_{in}^{MD}(Q_{A}+Q_{B})=Ne^{\frac{{(Q_{A}+Q_{B})^{2}}}{2\Delta^{2}}}\label{initsum}
\end{equation}
and the final state
\begin{equation}
\Psi_{fin}^{MD}(Q_{A}+Q_{B})=Ne^{\frac{{(Q_{A}+Q_{B}-2)^{2}}}{2\Delta^{2}}}\label{fini}
\end{equation}
This is a Gaussian around the weak value $(\sigma_{z}^{A}+\sigma_{z}^{B})_{w}=2$.
In fact for this example, the measurement need not be weak to get it right due to an accidental fact: the strong measurement also yields
the eigenvalue ``2'' with certainty.

Let us consider
another example to compare various measurement methods.
The system is described  by the two state vector
\begin{eqnarray}
\left\langle\uparrow\downarrow+\downarrow\uparrow+\uparrow\uparrow+\downarrow\downarrow\right|\;\;\;
\left|0.95\uparrow\downarrow-1.05\downarrow\uparrow+0.11\uparrow\uparrow\right\rangle
 \label{prepost1}
\end{eqnarray}
so that the weak value is
\begin{eqnarray}
(\sigma_{z}^{A}+\sigma_{z}^{B})_{w}= \frac{\left\langle
\uparrow\downarrow+\downarrow\uparrow+\uparrow\uparrow+\downarrow\downarrow\right|\sigma_{z}^{A}
+\sigma_{z}^{B}\left|-1.05\uparrow\downarrow+0.95\downarrow\uparrow+0.11\uparrow\uparrow\right\rangle
} {\left\langle
\uparrow\downarrow+\downarrow\uparrow+\uparrow\uparrow|-1.05\uparrow\downarrow+0.95\downarrow\uparrow+0.11\uparrow\uparrow\right\rangle}=22,
\end{eqnarray}
while the local weak values are
\begin{eqnarray}
& (\sigma_{z}^{A})_{w}=211\\
& (\sigma_{z}^{B})_{w}=-189.
\end{eqnarray}

Using the measuring device with the initial state (\ref{initsum}) we obtain the final state
\begin{eqnarray}
 {\Psi_{fin}^{MD}(Q_{A},Q_{B})=}
&N[-1.05e^{\frac{{-(Q_{A}+Q_{B})^2}}{2\Delta^{2}}}+0.95e^{\frac{{-(Q_{A}+Q_{B})^2}}{2\Delta^{2}}}+ 0.11e^{\frac{{-(Q_{A}+Q_{B}-2)^2}}{2\Delta^{2}}}],
\end{eqnarray}
and the expectation value of the pointer variable:
 \begin{eqnarray}
 \langle
Q_{A}+Q_{B}\rangle=\frac{22.0\left(11.0{e^{2.0/{\Delta}^{2}}}-10.0\right)}{221.0{e^{2.0/{\Delta}^{2}}}-220.0}\approx22-\frac{2360}{\Delta^{2}}.
\end{eqnarray}

\begin{figure}[h]
\includegraphics[width=14pc]{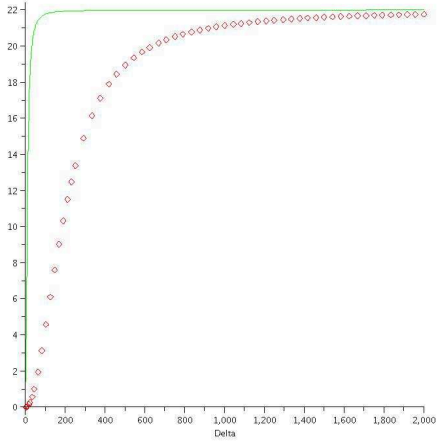}\hspace{2pc}%
\begin{minipage}[b]{14pc}\caption{\label{local vs nonlocal} {\bf Comparison of local and entangled measuring
devices}. Expectation values of the pointer variable  as a function of $\Delta$. Deviations of 10\%   are at values
of $\Delta>30$ for the nonlocal measuring device ( full line) and
$\Delta>600$ for the local one (dots) . While Deviations of 1\% are
at values of $\Delta>100$ for the nonlocal measuring device and $\Delta>2,000$
for the local one.}\end{minipage}
\end{figure}

A deviation of 1\% and an uncertainty of 10\% will require an
ensemble of about $2.2\times10^{3}$  particles see fig \ref{local vs nonlocal}.
If on the other hand we have only local weak measurements, i.e. two local
measuring devices, we will get an expectation value of $\langle Q_{A}+Q_{B}\rangle\approx22-\frac{8.8\times10^{5}}{\Delta^{2}}$.
Here a deviation of 1\% and an uncertainty of 10 \% will require an
ensemble of about $8.2\times10^{5}$ particles. In this example we can see that entanglement in the measuring device provides an improvement of more than two orders of magnitude.


\section{LOCAL AND NONLOCAL WEAK MEASUREMENTS OF THE PRODUCT OF LOCAL VARIABLES}

We now have enough background to analyze recent results about the measurements of the product of separate local variables named {}``joint
weak values'' \cite{ReschSteinberg}. If we are given nonlocal (unphysical) interactions,
then a weak measurement of the product is not different from any other
weak measurement and all theory of local weak measurements is applicable.
However, if we consider only local interactions, the situation is
very different. The product rule does not hold, not only for expectation
values of strong measurements

\begin{equation}
\langle AB\rangle_{\Phi\Psi}\neq\langle A\rangle_{\Phi\Psi}\langle B\rangle_{\Phi\Psi}\label{strongprod}
\end{equation}
 but also for weak values:
\begin{equation}
(AB)_{w}\neq A_{w}B_{w}.\label{weakprod}
\end{equation}

At first glance it seems that local weak measurements cannot help us
find the weak value of the product. Here is an example of two pre-
and post-selected states which yield different values, of $(\sigma_{z}^{A}\sigma_{z}^{B})_{w}$,
but have the same \textit{joint} probability for the pointer variables of local weak
measurements. In both cases the initial state is the product
$|\uparrow_{x}\rangle_{A}|\uparrow_{x}\rangle_{B}$, but the example
is more transparent if we write it in spin $z$ basis
\begin{equation}
|\Psi\rangle=\frac{1}{2}(|\uparrow\rangle_{A}|\uparrow\rangle_{B}+|\downarrow\rangle_{A}|\downarrow\rangle_{B}+|\uparrow\rangle_{A}|\downarrow\rangle_{B}+|\downarrow\rangle_{A}|\uparrow\rangle_{B})
\label{insta3}\end{equation}
 The first post-selected stated state is
\begin{equation}
|\Phi\rangle=\frac{1}{2}[|\uparrow\rangle_{A}|\uparrow\rangle_{B}+|\downarrow\rangle_{A}|\downarrow\rangle_{B}+i(|\uparrow\rangle_{A}|\downarrow\rangle_{B}-|\downarrow\rangle_{A}|\uparrow\rangle_{B})]
\label{finsta3}\end{equation}
 and the second is
\begin{equation}
|\Phi'\rangle=\frac{1}{2}[|\uparrow\rangle_{A}|\uparrow\rangle_{B}-|\downarrow\rangle_{A}|\downarrow\rangle_{B}+i(|\uparrow\rangle_{A}|\downarrow\rangle_{B}+|\downarrow\rangle_{A}|\uparrow\rangle_{B})]
\label{finsta3'}\end{equation}
 It is easy to see that in the first case $(\sigma_{z}^{A}\sigma_{z}^{B})_{w}=1$
while in the second case $(\sigma_{z}^{A}\sigma_{z}^{B})_{w}=-1$. In fact,
it is a special case in which the weak value is equal to the result
which is obtained with certainty in a strong nonlocal measurement.

For local measurement, we use the same model as before and write down the state of the two local measuring devices as a product of Gaussians
\begin{equation}
\Psi_{in}^{MD}(Q_{A},Q_{B})=e^{-\frac{Q_{A}^{2}}{4\Delta^{2}}}e^{-\frac{Q_{B}^{2}}{4\Delta^{2}}}
\label{productMD}
\end{equation}
After the interaction and post-selection, the joint distribution
for the pointer variables $Q_{a},Q_{B}$ given by $\Psi^{\dagger}\Psi$ turns out to be the same in both cases,
so that a measurement of $Q_{A} ,Q_{B}$ or any combination of the two will not provide us with a
method for distinguishing between the two initial states.

Nevertheless, Resch and Steinberg showed that one can find the weak value
of the product by looking at the local measuring device. They proved
the following formula (for a measuring device initially centered around zero)
\begin{equation}
Re(AB)_{w}=2\langle Q_{A}Q_{B}\rangle-Re(A_{w}^{*}B_{w})
\label{Resch}\end{equation}
Resch and Laudeen provided another expression
\begin{equation}
Re(AB)_{w}=\langle Q_{A}Q_{B}\rangle-\frac{4\Delta^{4}}{\hbar^{2}}\langle P_{A}P_{B}\rangle \label{laudeen}
\end{equation}
Now it is clear why there is no contradiction between the fact that
the probability distribution of the pointer variables of the measuring
devices are identical, while the joint weak values of the product are different.
It is not enough to look at the pointer variables, we have to look
at their conjugate momenta as well (a different type of measurement).
It is explicit in formula (\ref{laudeen}). (In formula (\ref{Resch}) we have the complex local weak variables and in order to see their imaginary parts we need to observe the conjugated momenta of the pointer variables.) 

An error analysis of this method for the example above shows that
 a set of measurements resulting in a deviation of$< 1\%$ and a statistical error of 10\%
requires an ensemble of about $2\times10^{6}$ such pre and post selected systems.
\begin{figure}[h]
\includegraphics[width=14pc]{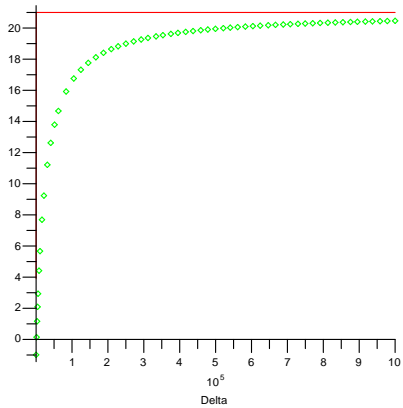}\hspace{2pc}%
\begin{minipage}[b]{14pc}\caption{\label{local vs nonlocal product}
{\bf Comparison of local and nonlocal (unphysical) measuring
devices for the measurement of a product}. Expectation values of the pointer variable  as a function of $\Delta$.
Deviations of 1\% are at values of $\Delta>100$ for the nonlocal
measuring device (full line) and $\Delta>5\times10^6$ for the local one
(dots) . }
\end{minipage}
\end{figure}
We obtain the weak value of the product from local weak measurements
but at the expense of extreme errors and with a requirement of two
different readings (one of Q and one of P).

For a comparison of this method and a non-local method with the (unphysical) coupling
term $g(t)\sigma_{z}^{A}\sigma_{z}^{B}P$ we look at the same pre and post-selected state as before (\ref{prepost1} ).
 The weak value for the product should be
\begin{equation}
A_{w}\equiv\frac{\langle{\Phi}\vert\sigma_{z}^{A}\sigma_{Z}^{B}\vert\Psi\rangle}{\langle{\Phi}\vert{\Psi}\rangle}=\frac{2-\epsilon^{2}}{2+\epsilon^{2}}=21\label{weakvalue4}
\end{equation}
As can be seen in fig \ref{local vs nonlocal product}, the non local
method converges much faster then the local one. A deviation of 1\%
with an uncertainty of 10\% would require an ensemble of about
$2\times10^{3}$ for the non-local method and about
$10^{12}$for the local method described above. It is not
surprising that this method is even less practical then the one for the
measurement of a sum.

\section{Conclusions}

Like eigenvalues and expectation values, weak values of a non local system can also be obtained using local methods. Such methods  require larger resources, and have no pointer pointing at the desired result.  Calculations using results from different types of measurement are required to arrive at the final result. We showed that the weak value of some non local variables (a sum of two or more local observables) can be measured directly. We have not found a direct way for weak measurement of a modular sum of nonlocal variables in spite of the existence of the method for strong measurement of nonlocal modular sum.

If we try to give an interpretation of  weak values as \textit{ elements of reality} \cite{elements}, one of the strengths of weak measurements is that it corresponds  to a shift of the pointer variable by the weak value. This is not the case when making local measurements for calculating non-local values. Such methods require us to look at different pointers  and use a formula for reaching the desired result.   Lundeen and Steinberg \cite{LundeenSteinberg09} measured non local weak values using the method of joint weak values to measure non local weak values in an optical experiment. Their results had large deviations (more than 25\%).  The question of interpretation still remains open but unlike local weak values which have been measured precisely  in the lab, some non-local weak values might still be thought of as accounting artifacts rather then physical observables.
\\

 This work has been supported in part by the European
Commission under the Integrated Project Qubit Applications (QAP)
funded by the IST directorate as Contract Number 015848 and by grant
990/06 of the Israel Science Foundation. We would like to thank Daniel Terno
and Judie Kupferman for their helpful comments.


\section*{References}

\bibliographystyle{unstr}

\end{document}